\RequirePackage{fix-cm}
\documentclass[smallextended]{svjour3}       
\smartqed  

\usepackage{amssymb}
\usepackage{amsmath}
\usepackage{braket}
\usepackage{booktabs}
\usepackage{graphicx}
\usepackage{subcaption}
\usepackage{caption}
\usepackage{placeins}
\usepackage{gensymb}
\usepackage{verbatim}
\usepackage[ruled,noend]{algorithm2e}
\usepackage{float}
\usepackage{xcolor}
\begin{document}

\title{Dimensionality Distinguishers
}


\author{Nayana Das         \and
       Goutam Paul 		   \and 
       Arpita Maitra
}

\institute{Nayana Das \at
              Applied Statistics Unit, Indian Statistical Institute, Kolkata 700108, India. \\
              \email{dasnayana92@gmail.com}           
           \and
           Goutam Paul \at
              Cryptology and Security Research Unit, R. C. Bose Centre for Cryptology and Security, Indian Statistical Institute, Kolkata 700108, India.\\
           \email{goutam.paul@isical.ac.in}
            \and  
           Arpita Maitra \at
              Indian Institute of Technology Kharagpur, Kharagpur 721302, India\\
		   \email{arpita76b@gmail.com}
}

\date{}

\maketitle

\begin{abstract}
The celebrated Clauser, Horne, Shimony and Holt (CHSH) game model helps to perform the security analysis of many two-player quantum protocols. This game specifies two Boolean functions whose outputs have to be computed to determine success or failure. It also specifies the measurement bases used by each player. In this paper, we generalize the CHSH game by considering all possible non-constant Boolean functions and all possible measurement basis (up to certain precision). Based on the success probability computation, we construct several equivalence classes and show how they can be used to generate three classes of dimension distinguishers. In particular, we demonstrate how to distinguish between dimensions 2 and 3 for a special form of maximally entangled state.
\end{abstract}

\section{Introduction}
\label{intro}
In quantum entanglement, two or more quantum particles (may be space-like separated) share their states in such a way that the state of each of the particles cannot be fully described without considering the other(s). If we change the quantum state of one particle thorough local unitary operations, the state of the rest of the particles changes automatically to maintain the entanglement. Many modern quantum protocols are based on entanglement theory. For example, quantum cryptography with Bell theorem~\cite{eke91}, super-dense coding~\cite{ben92}, quantum teleportation~\cite{ben93}, entanglement swapping~\cite{bose98} etc. Most of them use maximally entangled states. The Bell states are special cases of bipartite maximally entangled states on Hilbert space $\mathbb{C}^d\otimes\mathbb{C}^d$ given by $\ket{\psi} = U_A \otimes U_B \ket{\phi ^+_d}_{AB} $, where $\ket{\phi ^+_d}=\sum_{i=1}^{d}\frac{1}{\sqrt{d}}\ket{i}\otimes\ket{i}$ \cite{horo09}.

In 1935, Einstein, Podolsky and Rosen (EPR) showed that quantum mechanics is not complete \cite{eins35}. They also claimed that there may exist some local hidden variable theory, without requiring immediate action at a distance. Bell (1964) proposed a test for the existence of these hidden variables and developed an inequality \cite{bell64} and he showed that if the inequality is not satisfied, then a local hidden variable theory would not be possible. Inspired by Bell’s paper, Clauser, Horne, Shimony and Holt (CHSH) (1969) formed a correlation inequality and Bell's theorem can be proved by using that inequality \cite{chsh69}. The CHSH inequality gives a bound on any local hidden variable model (LHVM). \textcolor{black}{S. Cirel'son (1980) showed that Bell inequalities can be violated by quantum mechanical correlations \cite{cirel80}. A. Aspect, P. Grangier and G. Roger (1986) showed some experimental results, on the CHSH inequality, which agree with the quantum mechanical predictions \cite{gran86}. S. Popescu and D. Rohrlich (1994) formed some correlations, using no-signaling condition, violate the CHSH inequality even more than quantum  mechanical correlations \cite{pope94}. A simple setting for showing the usefulness of entanglement involves a two-player game known as the CHSH game \cite{handout,serg16}. Buhrman (2005) generalised the CHSH game in the field $F_q$ \cite{buhr05}. Some modern variants of CHSH appears in~\cite{tone08,pawl10,reic13,brun15,bava15}. } 

\subsection{Why dimensionality testing is important?}
	For a physical system, we generally assume that it has a particular dimension. Any practical application that uses entangled quantum systems have some predefined dimensional entangled states. In information theory, the dimensionality of quantum systems is a resource. In cryptographic applications, the security level scheme depends on the dimension. So testing dimensionality or distinguishing dimensionality of the underlying state-space are important pre-processing tasks before executing the actual protocol.
	
	Higher dimension implies more degrees of freedom. For example, consider  Quantum Key Distribution (QKD) protocol with qubit. In this case, the legitimate parties use only polarization of a photon for encoding. However, they have to fix the values for the other degrees of freedom such as spectral line, spatial mode or temporal mode etc. Lack of knowledge of any of these parameters may cause security back-door. Recently, Maitra et al.~\cite{arp18} showed that if the honest party measures only the polarization of a photon and remains ignorant about the {\em{Orbital Angular Momentum}} (OAM), then by changing the value of OAM one can steal more information than what he/she is entitled to in a certain type of QKD protocol. This strengthens the motivation of dimensionality testing.
	
\subsection{How to test the dimensionality?}
The dimension witness gives a bound on the dimension of an unknown system based on measurement statistics. It was first introduced for quantum systems in the context of non-local correlations by Brunner et al. \cite{bru08} and further developed in \cite{Pal08,Per08,Ver10,Ver08,Jun10,Bri11,weh08,gal10,jun1}. Various experiments have been recently proposed about the implementation of such witnesses \cite{joh12,mar12}. 	
	 
	Some theory of dimensional detection of an unknown quantum system is based on the set of conditional probabilities. It is based on the analysis on the probabilities of observing an outcome after creating and measuring the system for a given set of possibilities. It has become a prominent research area in recent times~\cite{weh08,gal10,jun1}. Experimental tests for testing dimension of a quantum system have been explored \cite{mar12,ahr12} and it has produced successful results.		
	A simple and general dimension witnesses for quantum systems of arbitrary Hilbert space dimension was proposed by Brunner (2013) \cite{bru13}. Their proposed work can distinguish between classical and quantum systems of the same dimension. A simple method for generating nonlinear dimension witnesses for systems of arbitrary dimension has been proposed by Bowles (2014) \cite{bow14}. It has been shown in this paper that this witness can be used to certify the presence of randomness. 

\subsection{Our contributions}
	In this paper, we generalize the CHSH game and define two classes of new games which are similar to the CHSH game. The first one is for $2$-variables and the second one is for $3$-variables. In this class of new games we change the winning condition of the CHSH game. Instead of a particular Boolean function in the CHSH game, we use all non-constant Boolean functions and find equivalence class for function pairs and bases such that, all the elements of the same class have the same winning probability of the game. We also consider all possible measurements subject to a precision parameter. For both the games, we optimize the winning probabilities. Finally, we show how our results can be used to devise three classes of dimensionality distinguishers, particularly between dimensions 2 and 3.

	The efficiency of a distinguisher depends on the number of samples (for a given success probability) and that in turn depends on the gap between the probabilities. This issue has been discussed in detail in~\cite{paul18}. Moreover, there are some works~\cite{basak18} on how to deal with finite number of samples. In the current work, we do not focus on these types of analysis. Rather, our main goal is to identify the distinguishing events with a significant probability gap and that is what we report here.

\section{Entanglement and the CHSH game}
A special type of entangled states are maximally entangled states. There are many quantum protocols which use these maximally entangled states. One of them is the CHSH game and we discuss about it.

\subsection{Maximally Entangled State}
Let us take a Hilbert space $H$ (for now, $H=\mathbb{C}^2$). There are infinitely many maximally entangled states in $H\times H$ and all are connected by a unitary. A pure bipartite state in $\mathbb{C}^2\times \mathbb{C}^2$ is maximally entangled if the reduced density matrix is $\frac{I}{2}$ for both sub systems.

Let $\ket{\phi}=\cos{\alpha}\ket{0}+\sin{\alpha}\ket{1}$ and $\ket{\theta}=\cos{\beta}\ket{0}+\sin{\beta}\ket{1}$.
Then $\ket{\Psi}_{AB}=\dfrac{1}{\sqrt{2}}[\ket{\phi\theta}+\ket{\phi^{\perp}\theta^{\perp}}]$
         $~=\dfrac{1}{\sqrt{2}}[\ket{0}\ket{\varphi}+ \ket{1}\ket{\varphi^{\perp}}]$, where $\ket{\varphi}=\cos({\alpha-\beta})\ket{0}-\sin({\alpha-\beta})\ket{1}$ 
is maximally entangled as $\rho_{A}=\rho_{B}=\dfrac{I}{2}$ where $\rho_{A}$ and $\rho_{B}$ are reduced density matrix of subsystem A and B respectively.

Again let $\ket{\chi}_{AB}=\dfrac{1}{\sqrt{2}}[\ket{0}\ket{\sigma}+\ket{1}\ket{\varrho}]$ where $\ket{\sigma}=\cos{\gamma}\ket{0}+\sin{\gamma}\ket{1}$ and $\ket{\varrho}=\cos{\delta}\ket{0}+\sin{\delta}\ket{1}$. To make $\rho_{A}=\rho_{B}=\dfrac{I}{2}$, we must have
$\ket{\varrho}=\ket{\sigma^{\perp}}$.

Thus a general form of maximally entangled state in $\mathbb{C}^2$ is $\dfrac{1}{\sqrt{2}}[\ket{\phi\theta}+\ket{\phi^{\perp}\theta^{\perp}}]$ (we are considering real coefficients only).

A maximally entangled (pure) state in a $\mbox{\textit{d}-dimensional}$ Hilbert space has the Schmidt decomposition $\sum_{i=1}^{d}\frac{1}{\sqrt{d}}\ket{i}\otimes\ket{i}$  in an appropriate basis. In Hilbert space $\mathbb{C}^m\otimes\mathbb{C}^n$ (say, $m<n$), a maximally entangled (pure) state is the same as that in $\mathbb{C}^m\otimes\mathbb{C}^m$.

\subsection{The CHSH Game}
\textcolor{black}{In this game there are two players, namely, Alice and Bob, and a referee.  Let us assume that Alice and Bob are far away from each other and not able to communicate during the game. Before the game begins, they can communicate freely to discuss their strategy. During the game, they only communicate with the referee in the following way:}\\
$\bullet$ The referee chooses two independent random bits $x$ and $y$ uniformly (also called “questions”) and sends $x$ to Alice and $y$ to Bob, 
\textcolor{black}{i.e., for all $s\in \{0,1\}, ~t\in \{0,1\}, ~\Pr(x=s,y=t)=\Pr_{xy}(s,t)=\dfrac{1}{4}$.}\\
$\bullet$ Alice and Bob reply to referee with bits $a$ and $b$ respectively.\\
$\bullet$ Referee calculates $x\wedge y$ and $a\oplus b$ (where $\wedge$, $\oplus$ stand for $AND$ and $XOR$ operations respectively)\\
$\bullet$ Alice and Bob win if $x\wedge y=a\oplus b$. 

Their goal is to achieve the highest winning probability together. 
Classically, the winning probability is $0.75$. But in the quantum world, this probability is $0.85$ if they follow the strategy discussed in the following subsection \ref{qstra}.

\subsubsection{Quantum Strategy}
\label{qstra}
\textcolor{black}{The strategy to win the game with maximum probability is to share a maximally entangled state (e.g, Bell state) between Alice and Bob. According to the referee's questions, they choose measurement bases to measure their qubits and send their answers to the referee. Details are given in the Algorithm~\ref{algo:game-1}. The values of $\theta_0$ and $\theta_1$ (defined in  Algorithm~\ref{algo:game-1}) are fixed for CHSH game and those values are $\theta_0=\dfrac{\pi}{8}$, $\theta_1=\dfrac{15\pi}{8}$.}


\textcolor{black}{
\subsubsection{Winning Probability}
Let $win$ be the event that Alice and Bob win, i.e., $x\wedge y=a\oplus b$. Now the winning probability of the CHSH game can be written as:\\
\begin{equation}\label{equ:win}
	\Pr(win)=\sum_{s,t}{\Pr}_{xy}(s,t)\Pr{(win|x=s,y=t)},
\end{equation}
which again implies that for $u,v,s,t \in \{0,1\}$,
$$\Pr(win)=\sum \limits_{s,t,u,v}{\Pr}_{xy}(s,t)(s\wedge t=u\oplus v){\Pr}_{ab|xy}(a=u,b=v|x=s,y=t).$$}

\textcolor{black}{
If the referee sends questions $x = 0, y = 0$, Alice and Bob win if they answer identically $a = 0, b = 0$ or $a = 1, b = 1$.}

\textcolor{black}{
	Then from Algorithm~\ref{algo:game-1}, the corresponding probability of winning (given $x = 0, y = 0$) is:\\
$\Pr(win|x=0,y=0) =
|\bra{0}\otimes\bra{\nu_0(\theta_{0})}\Psi_{AB}\rangle|^2 + |\bra{1}\otimes\bra{\nu_1(\theta_{0})}\Psi_{AB}\rangle|^2= \cos^{2}{\theta_0}$.}

\textcolor{black}{
Similarly we have,\\
$\Pr(win|x=0,y=1) =
|\bra{0}\otimes\bra{\nu_0(\theta_{1})}\Psi_{AB}\rangle|^2 + |\bra{1}\otimes\bra{\nu_1(\theta_{1})}\Psi_{AB}\rangle|^2= \cos^{2}{\theta_1}$,
\\
$\Pr(win|x=1,y=0) =
|\bra{0_{x}}\otimes\bra{\nu_0(\theta_{0})}\Psi_{AB}\rangle|^2 + |\bra{1_{x}}\otimes\bra{\nu_1(\theta_{0})}\Psi_{AB}\rangle|^2=\frac{1}{2}(1+\sin{2\theta_0})$,
\\
$\Pr(win|x=1,y=1) =
|\bra{0_{x}}\otimes\bra{\nu_1(\theta_{1})}\Psi_{AB}\rangle|^2 + |\bra{1_{x}}\otimes\bra{\nu_0(\theta_{1})}\Psi_{AB}\rangle|^2=\frac{1}{2}(1-\sin{2\theta_1})$.\\
}

\textcolor{black}{
Hence from Equation~\eqref{equ:win},\\
\begin{align*}
	P(win) &= \frac{1}{4}(P(win|x=0,y=0) + P(win|x=0,y=1)+ P(win|x=1,y=0)+ P(win|x=1,y=1))\\
 &=\frac{1}{4}[ \cos^{2}{\theta_0}+\cos^{2}{\theta_1}+\dfrac{1}{2}(1+\sin{2\theta_0})+\dfrac{1}{2}(1-\sin{2\theta_1})]. 
\end{align*}
}

\textcolor{black}{
This probability is maximum at $(\theta_0=\dfrac{\pi}{8}, \theta_1=\dfrac{15\pi}{8})$ and the maximum value is approximately $ 0.85355$. 
}

\section{Generalized Version of the CHSH Game}
We generalize the well known CHSH game to produce two types of new games. The first type of games are for $2$-variables (i.e., each question has $2$ options to answer). The other type of games are for $3$-variables (i.e., each question has $3$ options to answer).

\textcolor{black}{Here also we assume that Alice and Bob are far away from each other and not able to communicate during the game. Before the game begins, they can communicate freely to discuss their strategy. During the game, they only communicate with the referee.}

\subsection{New Games for $2$-variables (Game-1)}
Our new games are similar to the CHSH game. The only exception is in the winning condition.
Here the winning condition is $f(x,y)=g_2(a,b)$, where $f$ and $g_2$ are any two variable Boolean functions other than the constant functions (the subscript $2$ in $g_2$ is for $2$-variables). For $2$ variables, there are $(2^2)^2=16$ possible Boolean functions. Among them $2$ are constant functions. So we are playing this game with $14\times14=196$ pairs of function where in the CHSH game there is only one pair.

\textcolor{black}{
\subsubsection{Rules of Game-1}
For a fixed pair of two variable Boolean functions $(f,g_2)$ we define Game-1 as follows:\\
$\bullet$ The referee chooses two independent random bits $x$ and $y$ uniformly (also called “questions”) and sends $x$ to Alice and $y$ to Bob, i.e., for all $s\in \{0,1\}, ~t\in \{0,1\}, ~\Pr(x=s,y=t)=\Pr_{xy}(s,t)=\dfrac{1}{4}$.\\
$\bullet$ Alice and Bob reply to referee with bits $a$ and $b$ respectively.\\
$\bullet$ Referee calculates $f(x,y)$ and $g_2(a,b)$. \\
$\bullet$ Alice and Bob win if $f(x,y)=g_2(a,b)$. 
}

\textcolor{black}{
\subsubsection{Quantum Strategy for Game-1}
Alice and Bob follow the following strategy Algorithm~\ref{algo:game-1} to play Game-1. Here  also they share a maximally entangled state and choose measurement bases according to the referee's questions. They measure their qubits and send their answers to the referee. Alice's choice of measurement basis is only depends on referee's question. But for each pair $(f,g_2)$, Bob chooses the basis for which they can achieve maximum winning probability. Bob's bases are dependent on the parameters $\theta_0$ and $\theta_1$. So for different pairs of functions $(f,g_2)$, the values of $\theta_0$ and $\theta_1$ change. For example, CHSH game is a special case of Game-1, where $f=AND$, $g_2=XOR$, and Bob chooses $\theta_0=\frac{\pi}{8}$ and $\theta_1=\frac{15\pi}{8}$. } 

\begin{algorithm}[H]
	\caption{\textcolor{black}{Quantum Strategy for CHSH game and Game-1}}
	\label{algo:game-1}
	\begin{enumerate}
		\item \textcolor{black}{Before the game starts, Alice and Bob share $\ket{\Psi_{AB}} = \dfrac{1}{\sqrt{2}}(\ket{0}_A\otimes\ket{0}_B + \ket{1}_A\otimes\ket{1}_B)$}
		\item \textcolor{black}{Alice takes the first qubit and Bob takes the second qubit}
		\item \textbf{\textcolor{black}{Alice chooses:} }
		\begin{itemize}
			\item \textcolor{black}{Standard basis $\lbrace\ket{0},\ket{1}\rbrace$ if $x=0$
				\item Hadamard basis $\lbrace\ket{0_{x}},\ket{1_{x}}\rbrace$ if $x=1$, where \\
				$\ket{0_{x}}=\dfrac{1}{\sqrt{2}}(\ket{0}+\ket{1})$ and $\ket{1_{x}}=\dfrac{1}{\sqrt{2}}(\ket{0}-\ket{1})$}
		\end{itemize}					
		\item \textbf{\textcolor{black}{Bob chooses:}}\\
		\textcolor{black}{Basis $ \lbrace\ket{\nu_0(\theta_y)},\ket{\nu_1(\theta_y)}\rbrace$ corresponding to $y=0,1$,\\
			where $\ket{\nu_0(\theta_y)} = \cos \theta_y \ket{0} + \sin \theta_y \ket{1}$ , $\ket{\nu_1(\theta_y)} = \sin \theta_y\ket{0} - \cos \theta_y\ket{1}$, ~$0\leqslant \theta_0,\theta_1\leqslant 2\pi$}	
		\item \textbf{\textcolor{black}{Alice sends:}}
		\begin{itemize}
			\item \textcolor{black}{$a=0$ if  $\ket{0}$ or $\ket{0_{x}}$
				\item $a=1$ otherwise}
		\end{itemize}
		\item \textbf{\textcolor{black}{Bob sends:}}
		\begin{itemize}
			\item \textcolor{black}{$b=0$ if Bob gets $\ket{\nu_0(\theta_0)}$ or $\ket{\nu_0(\theta_1)}$
				\item $b=1$ otherwise}
		\end{itemize}
	\end{enumerate}
\end{algorithm}

\subsubsection{Success probabilities of Game-1}
We find the success probability of the game for each $f$ and $g_2$ by using Equation~\eqref{equ:win}, when the players follow the above strategy with changes in the chosen bases of Bob. Here Bob does not fix the value of $\theta_0$ and $\theta_1$. For different pairs of function $(f,g_2)$ the value of the pair $(\theta_0$ , $\theta_1)$ changes as the expression of the winning probability changes. 

For simplicity, we write an $n$-variable Boolean function as a $2^n$-length binary vector consisting of the last column of the truth table in lexicographical order, e.g., for a two variable function, we write $f(x,y)=[f(0,0),f(0,1), f(1,0),f(1,1)]$ and $g_2(a,b)=[g_2(0,0),g_2(0,1), g_2(1,0),g_2(1,1)]$. \textcolor{black}{Also LHS and RHS denote left hand side and right hand side respectively.}

The results are in the following Table~\ref{my-label}. \textcolor{black}{The first two columns of Table~\ref{my-label} represent the functions of inputs and outputs (i.e., $f (x, y)$ and $g_2(a, b)$ )respectively, and corresponding success probabilities are given in third column. The number of such function pair $(f, g_2)$ having same success probabilities are in the last column. }

\begin{table}[h!]
\centering
\caption{Success probabilities of Game-1 with any non-constant 2 variables Boolean functions $f$ and $g$}
\label{my-label}	
\resizebox{\columnwidth}{!}{
\begin{tabular}{|c|c|c|c|}
	\hline 
	\hline
	LHS of winning & RHS of winning & Success & Number of such \\ 
	condition $f (x, y)$ & condition $g_2(a, b)$ &  probability& function pair $(f, g_2)$\\
	\hline 
	\hline
	any non constant $f$ & XOR, XNOR & $0.85$ & $28$ \\ 
	\hline 
	$f(x,y)$ contains one $0$ & $g_2(a,b)$ contains one $0$ & $0.80$ & $32$ \\ 
	\hline 
	$f(x,y)$ contains one $1$ & $g_2(a,b)$ contains one $ 1$ & $0.80$ & $32$ \\ 
	\hline 
	$f(x,y)$ contains two $0$ & $g_2(a,b)$ contains either & $0.67$ & $48$ \\ 
	 & exactly one $1$ or $0$&  &  \\ 	
	\hline 
	$f(x,y)$ contains one $1$ & $g_2(a,b)$ contains one $0$ & $0.55$& $16$ \\ 
	\hline 
	$f(x,y)$ contains one $0$ & $g_2(a,b)$ contains one $1$ &$0.55$ & $6$ \\ 
	\hline 
	Any non-constant $f$ & $g_2(a,b)=a,~ b$, $\bar{a}$, $\bar{b}$ & $0.5$ & $56$ \\ 
	\hline 
	\hline
	\end{tabular}
	}
	\end{table}

\begin{figure}[h!]

  \centering
  \begin{subfigure}{0.49\textwidth}
    \includegraphics[width=1.0\linewidth]{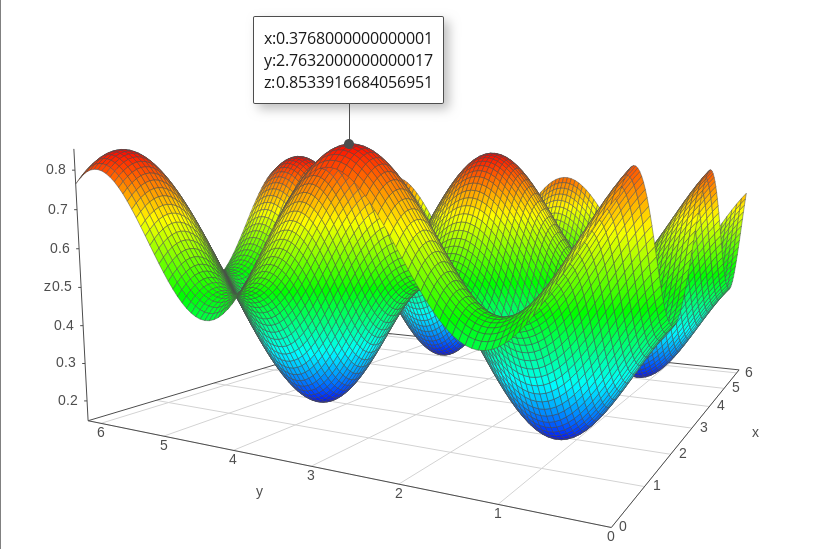}
     \caption{$f=AND,~ g_2=XOR.$}
  \end{subfigure}
  \begin{subfigure}{0.49\textwidth}
    \includegraphics[width=1.0\linewidth]{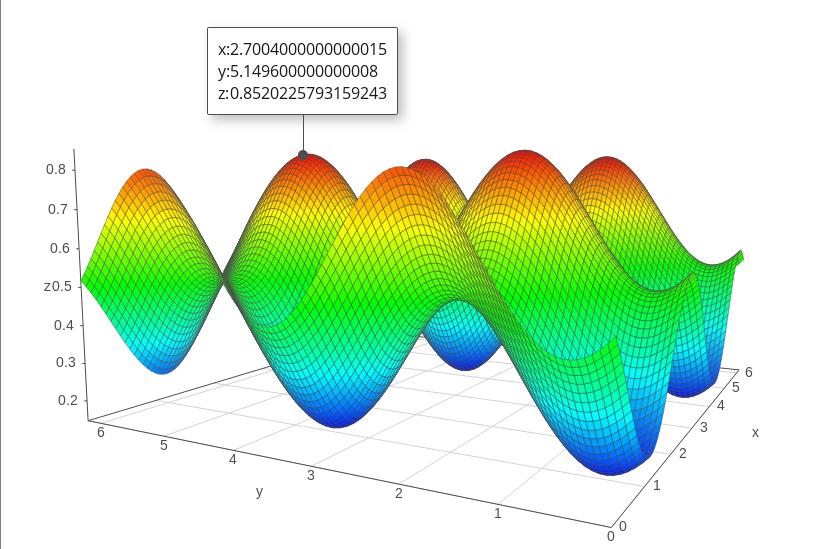}
    \caption{$f=OR,~ g_2=XOR.$}
  \end{subfigure}
  \begin{subfigure}{0.49\textwidth}
    \includegraphics[width=\linewidth]{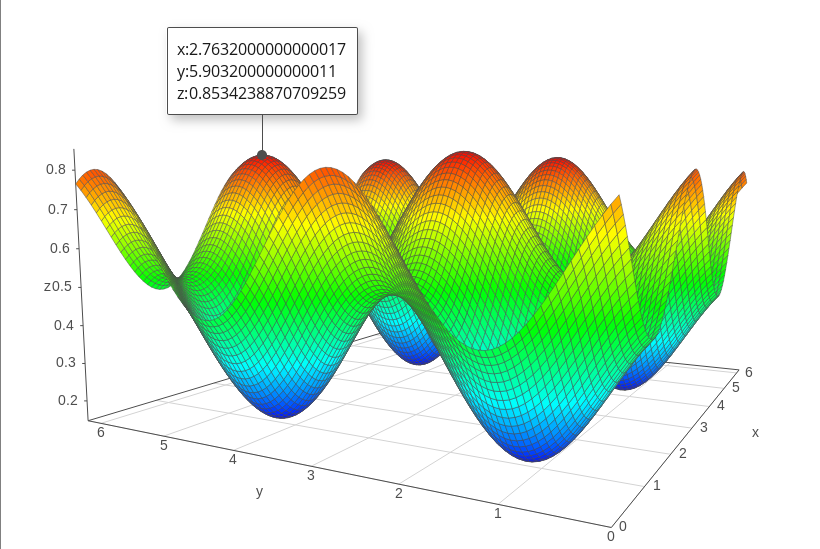}
    \caption{$f=x,~ g_2=XOR$.}
  \end{subfigure}
  \begin{subfigure}{0.49\textwidth}
    \includegraphics[width=\linewidth]{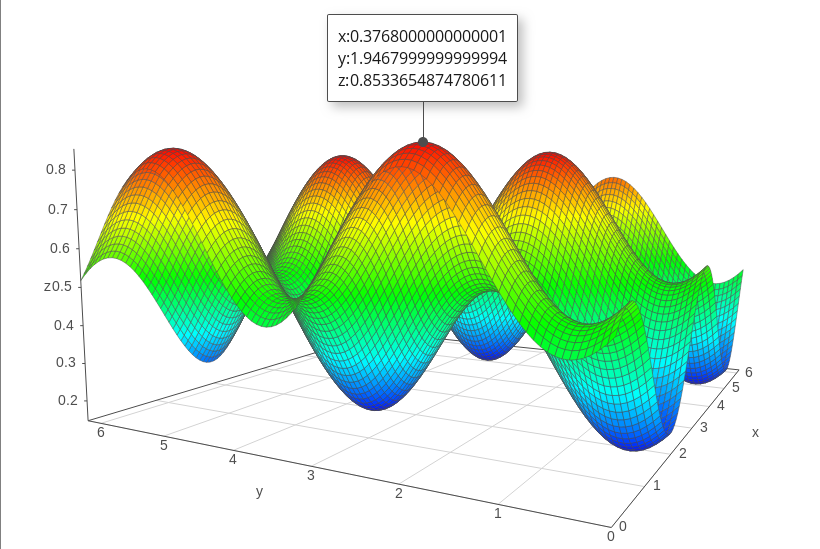}
    \caption{$f=y,~ g_2=XOR.$}
  \end{subfigure}
  \caption{Success probability graphs for 4 different cases of Game-1 with non-constant 2 variables Boolean functions $f$ and $g_2$.}
  \label{fig:success}
\end{figure}

\subsubsection{Observation}
From Table~\ref{my-label}, we observe that the winning probability is maximum when $g_2(a,b)=a\oplus b$ and $a\odot b$, i.e., for any non-constant $2$ variables Boolean function $f$, if $g_2=XOR$ or $g_2=XNOR$ then by playing the Game-1 we can win the game with probability $0.85$.

The reason behind this is that the probability graph of these 28 cases are almost similar. To illustrate this, we show some probability graphs in Figure~\ref{fig:success}. 
\textcolor{black}{In these graphs we plot $\theta_0$ ($x$-axis) vs. $\theta_1$($y$-axis) vs. success probability expression ($z$-axis). From these graphs we can see that for each case the success probabilities are periodic functions of $(\theta_0,\theta_1)$ and achieve maximum value $0.85$ at more than one points.}

\textcolor{black}{$\bullet$ The first graph in Figure~\ref{fig:success}(a) represents the success probability $\frac{1}{4}[1+ cos^2 \theta_0+ cos^2 \theta_1+\frac{1}{2}sin{2\theta_0}-\frac{1}{2}sin{2\theta_1}]$ corresponding to the function pair $(f=AND,~ g_2=XOR)$ and one of its maximum point is at $\left(\theta_0=\dfrac{\pi}{8},\theta_1=\dfrac{15\pi}{8}\right)$.
}

\textcolor{black}{$\bullet$ The second graph in Figure~\ref{fig:success}(b) represents the success probability $\frac{1}{4}[1+ cos^2 \theta_0+ sin^2 \theta_1-\frac{1}{2}sin{2\theta_0}-\frac{1}{2}sin{2\theta_1}]$ corresponding to the function pair $(f=OR,~ g_2=XOR)$ and one of its maximum point is at $\left(\theta_0=\dfrac{7\pi}{8},\theta_1=\dfrac{5\pi}{8}\right)$.
}

\textcolor{black}{$\bullet$ The third graph in Figure~\ref{fig:success}(c) represents the success probability $\frac{1}{4}[1+ cos^2 \theta_0+ cos^2 \theta_1-\frac{1}{2}sin{2\theta_0}-\frac{1}{2}sin{2\theta_1}]$ corresponding to the function pair $(f=x,~ g_2=XOR)$, where $f=x$ means $f(x,y)=x ~\forall~ x,y \in \{0,1\}$, and one of its maximum point is at $\left(\theta_0=\dfrac{7\pi}{8},\theta_1=\dfrac{7\pi}{8}\right)$.
}

\textcolor{black}{$\bullet$ The fourth graph in Figure~\ref{fig:success}(d) represents the success probability $\frac{1}{4}[1+ cos^2 \theta_0+ sin^2 \theta_1+\frac{1}{2}sin{2\theta_0}-\frac{1}{2}sin{2\theta_1}]$ corresponding to the function pair $(f=y,~ g_2=XOR)$, where $f=y$ means $f(x,y)=y ~\forall~ x,y \in \{0,1\}$, and one of its maximum point is at $\left(\theta_0=\dfrac{9\pi}{8},\theta_1=\dfrac{5\pi}{8}\right)$.
}

\subsection{New Games for $3$-variables (Game-2)}

In this game there are two players, namely, Alice and Bob (they are far away  from each other and not able to communicate) and a referee. Let us define the sets $S=\lbrace0,1,2\rbrace$, $\mathcal{G}=\lbrace g:S\times S\rightarrow \lbrace0,1\rbrace\rbrace$ and $\mathcal{F}=\{f:f~ is~ a~ 2~ variable~ Boolean function\}$.\\

\textcolor{black}{
\subsubsection{Rules of Game-2}
For a particular pair $(f,g_3)$, where $f\in\mathcal{F}$ and $g_3\in\mathcal{G}$ (the subscript $3$ in $g_3$ is for $3$-variables), we define Game-2 as follows:\\
$\bullet$ The referee chooses two independent random bits $x$ and $y$ uniformly (also called “questions”) and sends $x$ to Alice and $y$ to Bob.
That is, for all $s\in \{0,1\}, ~t\in \{0,1\}, ~Pr(x=s,y=t)=P_{xy}(s,t)=\dfrac{1}{4}$.\\
$\bullet$ Alice and Bob send their answers $a$ and $b$ ($a,b\in\{0,1,2\}$) to the referee.\\
$\bullet$ Referee calculates $f(x,y)$ and $g_3(a,b)$. \\
$\bullet$ Alice and Bob win if $f(x,y)=g_3(a,b)$. 
}

\textcolor{black}{
\subsubsection{Quantum Strategy for Game-2}
Now let Alice and Bob play the game with the following strategy given in Algorithm~\ref{algo:game-2}. Before the game starts, they share a maximally entangled bipartite state: $\ket{\Psi_{AB}} = \frac{1}{\sqrt{3}}(\ket{0}_A\otimes\ket{0}_B + \ket{1}_A\otimes\ket{1}_B + \ket{2}_A\otimes\ket{2}_B)$ in the Hilbert space $\mathbb{C}^3\otimes \mathbb{C}^3$. According to the referee's questions, they choose measurement bases to measure their qubits and send their answers to the referee. Alice's choice of measurement basis is only depends on referee's question. But for each pair $(f,g_3)$, Bob choose the basis for which they can achieve maximum winning probability. Bob's bases are dependent on the parameters $\theta_0$ and $\theta_1$. 
 }

\begin{algorithm}[H]
	\caption{\textcolor{black}{Quantum Strategy for Game-2}}
	\label{algo:game-2}
	\begin{enumerate}
		\item \textcolor{black}{Before the game starts, Alice and Bob share\\
			$\ket{\Psi_{AB}} = \dfrac{1}{\sqrt{3}}(\ket{0}_A\otimes\ket{0}_B + \ket{1}_A\otimes\ket{1}_B + \ket{2}_A\otimes\ket{2}_B)$ }
			\item \textcolor{black}{Alice takes the first qubit and Bob takes the second qubit}
			\item \textcolor{black}{\textbf{Alice chooses:}}
		\begin{itemize}
			\item \textcolor{black}{Standard basis $\lbrace\ket{0},\ket{1},\ket{2}\rbrace$ if $x=0$ 
				\item Fourier basis $\lbrace\ket{0_{x}},\ket{1_{x}},\ket{2_{x}}\rbrace$ if $x=1$, where\\
				$\ket{0_{x}}=\dfrac{1}{\sqrt{3}}(\ket{0}+\ket{1}+\ket{2})$, $\ket{1_{x}}=\dfrac{1}{\sqrt{3}}(\ket{0}+\omega\ket{1} +\omega^{2}\ket{2})$,\\
				$\ket{2_{x}}=\dfrac{1}{\sqrt{3}}(\ket{0}+\omega^{2}\ket{1} +\omega\ket{2})$ and $\omega=e^{{2\pi i}/{3}}$}
		\end{itemize}
		\item \textbf{\textcolor{black}{Bob chooses:}}
		\begin{itemize}
			\item \textcolor{black}{Basis $ \lbrace\ket{\psi_0},\ket{\psi_1},\ket{\psi_2}\rbrace$ if $y=0$, \\
				$\ket{\psi_0} = \cos \theta_0 \ket{0} + \sin \theta_0  \cos \theta_1 \ket{1} + \sin \theta_0 \sin \theta_1 \ket{2}$\\
				$\ket{\psi_1} = \sin \theta_0 \ket{0} - \cos \theta_0  \cos \theta_1 \ket{1} - \cos \theta_0 \sin \theta_1 \ket{2}$\\
				$\ket{\psi_2} = \sin \theta_1 \ket{1} +  \cos \theta_1 \ket{2}$ and\\ 
				$0\leqslant \theta_0,\theta_1\leqslant 2\pi$
				\item Basis $ \lbrace\ket{\phi_0},\ket{\phi_1},\ket{\phi_2}\rbrace$ if $y=1$,\\
				$\ket{\phi_0} = \cos \theta_1 \ket{0} + \sin \theta_1 \cos \theta_0 \ket{1} + \sin \theta_1 \sin \theta_0 \ket{2}$\\
				$\ket{\phi_1} = \sin \theta_1 \ket{0} - \cos \theta_1 \cos \theta_0 \ket{1} - \cos \theta_1 \sin \theta_0 \ket{2}$ \\
				$\ket{\phi_2} = \sin \theta_0 \ket{1} +  \cos \theta_0 \ket{2}$ and\\
				$0\leqslant \theta_0,\theta_1\leqslant 2\pi$}
		\end{itemize}
		
		\item \textbf{\textcolor{black}{Alice sends:}}
		\begin{itemize}
			\item \textcolor{black}{$a=0$ if Alice gets $\ket{0}$ or $\ket{0_{x}}$
				\item $a=1$ if she gets $\ket{1}$ or $\ket{1_{x}}$
				\item $a=2$ otherwise}
		\end{itemize}
		
		\item \textbf{\textcolor{black}{Bob sends:}}
		\begin{itemize}
			\item \textcolor{black}{$b=0$ if Bob gets $\ket{\psi_0}$ or $\ket{\phi_0}$
				\item $b=1$ if he gets $\ket{\psi_1}$ or $\ket{\phi_1}$ 
				\item $b=2$ otherwise}
		\end{itemize}
	\end{enumerate}
\end{algorithm}

\subsubsection{Example of Game-2}
Let us take an example. Let $f(x,y)=x\wedge y$ and $g_3(a,b)=a~Embedded~XOR~b$ (i.e., $g_3(a,b)=0~if~a=b ~and~g_3(a,b)=1~otherwise$).
If we play the above game with these $f$ and $g_3$ then the success probability is $0.76$ at $\theta_0= \dfrac{17\pi}{16},\theta_1=\dfrac{\pi}{16}$.

\subsubsection{Maximum Winning Probability}
In this Game-2 the maximum winning probability is $0.86$ only for $8$ pairs of function $(f,g_3)$. 

Now the function pairs, with the highest winning probability and corresponding bases are shown in Table~\ref{Max-game2}.

\begin{table}[h]
\centering
\caption{Function pairs with maximum success probabilities of Game-2}
\label{Max-game2}	
\begin{tabular}{|c|c|c|c|}
	\hline 
	\hline
	$f$ & $g_3$ & $\theta_0$ & $\theta_1$  \\
	\hline 
	\hline
	[0, 1, 0, 0] &[0, 1, 0, 1, 0, 0, 0, 0, 1] & ${33\pi}/{32}$ & ${19\pi}/{32}$ \\ 
	\hline 
	[0, 1, 0, 0] & [1, 0, 0, 0, 0, 1, 0, 1, 0] & ${	29\pi}/{32}$ & ${29\pi}/{32}$ \\ 
	\hline 
	[0, 1, 1, 1] & [0, 1, 1, 1, 1, 0, 1, 0, 1] & ${	29\pi}/{32}$ & ${15\pi}/{32}$\\ 
	\hline 
	[0, 1, 1, 1] & [1, 0, 1, 0, 1, 1, 1, 1, 0] &	${	19\pi}/{32}$ & ${33\pi}/{32}	$\\ 
	\hline 
	[1, 0, 0, 0] & [0, 1, 0, 1, 0, 0, 0, 0, 1] &	${	19\pi}/{32}$ & ${33\pi}/{32}$	\\ 
	\hline 
	[1, 0, 0, 0] & [1, 0, 0, 0, 0, 1, 0, 1, 0] & ${	29\pi}/{32}$ & ${15\pi}/{32}$\\ 
	\hline 
	[1, 0, 1, 1] & [0, 1, 1, 1, 1, 0, 1, 0, 1] &	${	15\pi}/{32}$ & ${29\pi}/{32}$\\ 
	\hline 	
	[1, 0, 1, 1] & [1, 0, 1, 0, 1, 1, 1, 1, 0] &	${	33\pi}/{32}$ & ${19\pi}/{32}$\\
	\hline
	\hline 
	\end{tabular}
\end{table}
\subsection{Equivalence Classes}
From the results of these two games we observe that, if we introduce some equivalence relations to make partition of the set of data in each game result, then we will take only one element of each equivalence class to play these games. It will reduce the time and space complexity of these games. Also if some measurement setup will be unavailable then we can use any other setup from the same class to continue the games. Here we take three equivalence relations to make three different types of partitions of the results.

\begin{enumerate}
\item We can make an equivalence class of the bases of Bob for a fixed function pair $(f,g_i)$, ($i=2,3$), such that all elements of the same class give the same success probability.

For simplicity, we only write the value of the pair $(\theta_{1},\theta_{2})$ as a basis (i.e., we represent a basis as a point $(\theta_{1},\theta_{2})$ in $\mathbb{R}^2$ ) in a class and we take the values in $radian$ (i.e., $0\leqslant \theta_0,\theta_1\leqslant 2\pi$) and as a multiple of $\dfrac{\pi}{32}$.

For example, if we fix $f=AND$ and $g_2=XOR$ in Game-1, then there are $8$ equivalence classes of bases (up to $1$ significant digit). Now in the previous example, if we consider the success probabilities up to $2$ significant digits, then there are $4$ elements in the class of highest winning probability $0.85$ and the class is
\begin{align*}
 \left\{\left(\dfrac{\pi}{8},\dfrac{7\pi}{8}\right),\, \left(\dfrac{\pi}{8},\dfrac{15\pi}{8} \right),\, \left(\dfrac{9\pi}{8},\, \dfrac{7\pi}{8}\right),\, \left(\dfrac{9\pi}{8},\dfrac{15\pi}{8} \right)\right\}.
\end{align*}

Again in Game-2, let $f=AND$ and $g_3=Embedded~XOR$ (i.e. $g_3(a,b)=0~if~a=b ~and~g_3(a,b)=1~otherwise$) , then there are $7$ equivalence classes of bases (up to $1$ significant digit). Now in the previous example, if we consider the success probabilities up to $2$ significant digits, then there are $4$ elements in the class of highest winning probability $0.76$ and the class is 
\begin{align*}\left\{\left( \dfrac{33\pi}{32},\dfrac{\pi}{32}
\right),\left( \dfrac{33 \pi}{32},\dfrac{2\pi}{32}
\right), \left( \dfrac{34\pi}{32},\dfrac{\pi}{32}
\right),
 \left( \dfrac{34\pi}{32},\dfrac{2\pi}{32}
\right)\right\}.
\end{align*}
 
\item Secondly, we fix the bases of Bob and vary the function pairs to make the equivalence classes. Here also all the elements of the same class have the same  winning probability.

For example, in Game-1, if we fix $\left(\theta_0=\dfrac{\pi}{8}, \theta_1=\dfrac{15\pi}{8}\right)$, then $(f=[0, 0, 0, 1],~g_2=[0, 1, 0, 1] )$, $(f=[0, 0, 1, 0],~g_2=[0, 1, 0, 1])$, $(f=[0, 0, 1, 1],~g_2=[0, 1, 0, 1])$, $(f=[0, 1, 1, 1],~g_2=[0, 1, 0, 1])$ etc. are all belong to the same class with success probability $0.5$.

\item At last, we vary both function pairs and Bob's bases and the tuples which have the same winning probability are belong to the same class.
E.g., in Game-2, each row of Table~\ref{Max-game2} have the same success probability $0.86$ and thus they belong to the same class.
\end{enumerate}

\section{Dimensionality Testing}
We observe the winning probabilities of various cases in Game-1 and Game-2.

By using the above two games we can make device independent dimension distinguisher to distinguish between the states $\ket{\Psi_{AB}} = \dfrac{1}{\sqrt{2}}(\ket{0}_A\otimes\ket{0}_B + \ket{1}_A\otimes\ket{1}_B)$ and $\ket{\Phi_{AB}} = \dfrac{1}{\sqrt{3}}(\ket{0}_A\otimes\ket{0}_B + \ket{1}_A\otimes\ket{1}_B + \ket{2}_A\otimes\ket{2}_B)$. For example,

$\bullet$ In Game-1, if we take $f(x,y)=x\wedge y$ and $g_2(a,b)=a\oplus b$ and  $\theta_0=\dfrac{\pi}{8}$, $\theta_1=\dfrac{15\pi}{8}$, then the winning probability of this game is $0.85$.

$\bullet$ In Game-2, if we take $f(x,y)=x\wedge y$ and $g_3=Embedded~XOR$ and  $\theta_0=\dfrac{\pi}{8}$, $\theta_1=\dfrac{15\pi}{8}$, then winning probability of this game is $0.76$.

So by playing these games and observing winning probabilities we can easily distinguish between $\ket{\Psi_{AB}}$ and $\ket{\Phi_{AB}}$. In other words, we can say the dimension of the given maximally state is two or three.

We can think this whole process as a union of two black boxes. An initial black box is the state preparatory which prepares states of form either $\ket{\Psi_{AB}}$ or $\ket{\Phi_{AB}}$. the prepared state is then sent to a second black box, the measurement device. In this box, if the states are $\ket{\Psi_{AB}}$, it will follow the process of Game-1 and if the states are $\ket{\Phi_{AB}}$, it will follow the process of Game-2. 
	 
From the outputs of this measurement device we will calculate the winning probability of the game played in this box and compare this probability with the success probabilities of Game-1 and Game-2. So we have a dimension distinguisher. The protocol is described in Algorithm~\ref{algo:dim_dist}.

Following the above process and by changing the function pairs in the games we can find many distinguishers. For each, we use the function pair $(f,g_3)$ in Game-2 and the function pair $(f,g'_{2})$ in Game-1 (where, $g'_{2}$ is the restriction of $g_3$ in $2$ variables, i.e., $g'_{2}(a,b)=[g_3(0,0),g_3(0,1), g_3(1,0),g_3(1,1)]$ ). We divide the set of all distinguisher into $3$ classes according to the winning probabilities of the games.

\subsection{First class of distinguishers ($D_1$)}
	\textcolor{black}{In this set, we put all the distinguishers where we choose function pairs $(f,g_3)$ such that the function pair $(f,g'_{2})$ has the highest winning probability in Game-1 (i.e., $0.85$) which  is greater than the winning probability of the corresponding Game-2.}

	\textcolor{black}{If we choose $f=[0, 1, 0, 0]	,g_3=[0, 1, 1, 1, 0, 1, 1, 1, 0]$, thus $g'_{2}=[0, 1, 1, 0]$ (or $f=[0, 1, 1, 1],	g_3=[1, 0, 0, 0, 1, 0, 0, 0, 1]$, thus $g'_{2}=[0, 1, 1, 0]$), then the winning probabilities of the Game-1 and Game-2 are $0.85$ and $0.58$. Therefore the difference of these probabilities is $0.27$, which is quite good.}
	
	\textcolor{black}{There are many distinguishers in this class. We put some of them into the following Table~\ref{table D1}. Here we take the winning probability for $d=3$ at that point where the corresponding winning probability for $d=2$ is maximum.}

\begin{table}[h]
\caption{Table for $D_1$}
\label{table D1}	
\resizebox{\columnwidth}{!}{
\begin{tabular}{|c|c|c|c|c|c|}
\hline
\hline 
$f$ & $g'_{2}$ &  $g_3$ &  W.P. if $d=2$ & W.P. if $d=3$ & Difference \\ 
\hline 
\hline
[0, 0, 0, 1]& [0, 1, 1, 0]& 	 [0, 1, 0, 1, 0, 0, 0, 1, 1]& 	 0.85	& 	0.53	&  0.32\\
\hline
[0, 0, 0, 1]& [0, 1, 1, 0]& 	 [0, 1, 0, 1, 0, 0, 1, 1, 1]	&   0.85	&  	0.51 &  	0.34 \\
\hline
[0, 0, 0, 1]& [1, 0, 0, 1]&  [1, 0, 1, 0, 1, 1, 1, 1, 1]	&   0.85		&  0.45&  	0.4\\
\hline
[0, 0, 1, 0]& [1, 0, 0, 1]& 	 [1, 0, 0, 0, 1, 1, 1, 0, 1]	&   0.85	&  	0.41&  	0.44\\
\hline
[0, 0, 1, 1]&  [0, 1, 1, 0]&	 [0, 1, 0, 1, 0, 0, 0, 0, 1] &  0.85	&  	0.39&  	0.46 \\
\hline
[0, 1, 0, 0]&  [0, 1, 1, 0]& [0, 1, 1, 1, 0, 1, 1, 1, 1] &  0.85	& 	0.42	& 0.43\\
\hline
[0, 1, 0, 1]& [0, 1, 1, 0]&  [0, 1, 1, 1, 0, 1, 0, 1, 0]& 	 0.85	&  	0.46&  0.39\\
\hline
[0, 1, 1, 1]& [0, 1, 1, 0]& [0, 1, 0, 1, 0, 1, 0, 1, 0] &   0.85	&  	0.45&  	0.4\\
\hline 
[1, 0, 0, 1]&  [1, 0, 0, 1]& [1, 0, 1, 0, 1, 0, 1, 0, 1]	&  0.85	& 	0.53&  0.32\\
\hline
[1, 0, 1, 0]& [1, 0, 0, 1]&  [1, 0, 0, 0, 1, 0, 1, 0, 1]&  	0.85&  	0.46&  0.39\\
\hline
[1, 0, 1, 1]&  [0, 1, 1, 0]&  [0, 1, 0, 1, 0, 1, 0, 0, 0]	&   0.85	&  	0.44&  	0.41\\
\hline
[1, 1, 0, 0]&  [1, 0, 0, 1]&  [1, 0, 1, 0, 1, 1, 1, 1, 0]	&   0.85	&  	0.39&  	0.46\\
\hline
[1, 1, 1, 0]& [0, 1, 1, 0]& 	[0, 1, 1, 1, 0, 0, 0, 0, 0]	&  	 0.85	& 	0.41&  	0.44\\
\hline
\hline
\end{tabular}
}\\
\textit{\begin{tiny}
*W.P denotes winning probability.
\end{tiny}}

\end{table}

\subsection{Second class of distinguishers ($D_2$)}
	\textcolor{black}{In this set, we put all the distinguishers where we choose function pairs $(f,g_3)$ such that it has the highest winning probability in Game-2 (i.e., $0.86$) which  is greater than the winning probability of corresponding Game-1 with function pair $(f,g'_{2})$. Here we take the winning probability for $d=2$ at that point where the corresponding winning probability for $d=3$ is maximum.}

	\textcolor{black}{For example, let $f=[0, 1, 0, 0],	g_3=[1, 0, 0, 0, 0, 1, 0, 1, 0]$ then if $d=2$ success probability is $0.80$ and if $d=3$ success probability is $0.86$. We put all distinguishers in Table~\ref{table D2}.}

\begin{table}[ht]
\caption{Table for $D_2$}
\label{table D2}
\resizebox{\columnwidth}{!}{
\begin{tabular}{|c|c|c|c|c|c|}
\hline
\hline 
 $f$ &  $g'_{2}$ &   $g_3$ &   W.P. if $d=2$ &  W.P. if $d=3$ &   Difference  \\ 
\hline 
\hline
[0, 1, 0, 0] & [0, 1, 1, 0] & [0, 1, 0, 1, 0, 0, 0, 0, 1] & 0.46 & 0.86 &  0.4\\ 
\hline 
[0, 1, 0, 0] & [1, 0, 0, 0] & [1, 0, 0, 0, 0, 1, 0, 1, 0] & 0.64	& 	0.86 & 	0.22   \\ 
\hline 
[0, 1, 1, 1]& [0, 1, 1, 1]& [0, 1, 1, 1, 1, 0, 1, 0, 1]& 	 0.63	& 	0.86 & 	0.23 \\ 
\hline 
[0, 1, 1, 1] & [1, 0, 0, 1]	& [1, 0, 1, 0, 1, 1, 1, 1, 0]	 &   0.48 & 	0.86	 &  0.38 \\ 
\hline 
[1, 0, 0, 0] & [0, 1, 1, 0] &  [0, 1, 0, 1, 0, 0, 0, 0, 1]	 & 0.48	 & 0.86 &  	0.38 \\ 
\hline 
[1, 0, 0, 0] &  [1, 0, 0, 0] & [1, 0, 0, 0, 0, 1, 0, 1, 0] & 	 0.63 & 	0.86 & 	0.23\\ 
\hline 
[1, 0, 1, 1] &  [0, 1, 1, 1] & [0, 1, 1, 1, 1, 0, 1, 0, 1] &  0.64& 	0.86 & 	0.22  \\ 
\hline 
[1, 0, 1, 1]& [1, 0, 0, 1]& 	 [1, 0, 1, 0, 1, 1, 1, 1, 0] &	 0.46	& 	0.86&  	0.4 \\ 
\hline 
\hline
\end{tabular}
}\\
\textit{\begin{tiny}
*W.P denotes winning probability.
\end{tiny}}
 
\end{table}

\subsection{Third class of distinguishers ($D_3$)}
	\textcolor{black}{Similarly, we can make dimension distinguisher using other function pairs for which the difference between the optimal winning probabilities of the two games is non-negligible. Here we take function pair $(f,g_3)$ and corresponding pair $(f,g'_{2})$ such that both the games with respective pairs do not achieve the highest winning probabilities. we put all these distinguishers in this set. The cardinality of this set depends on the difference value between winning probabilities. }
	
	\textcolor{black}{Let $(f,g_3)$ be a function pair and the highest winning probability of Game-2 with $(f,g_3)$ being $p_2$ at point $(s_2,t_2)$ and the same of Game-1 with $(f,g'_{2})$ is $p_1$ at point $(s_1,t_1)$. We compare $p_1$, $p_2$ and take the best (say, $p_1>p_2$). Then we find the 
winning probability $p$ of Game-2 at $(s_1,t_1)$ and difference value $p_1-p$. We make a list of these distinguishers for which the difference value is greater than $0.44$ in Table~\ref{table D3}.}

\begin{table}[h]
\caption{Table for $D_3$}
\label{table D3}
\resizebox{\columnwidth}{!}{
\begin{tabular}{|c|c|c|c|c|c|}
\hline
\hline 
$f$ & $g'_{2}$ &  $g_3$ &  W.P. if $d=2$ & W.P. if $d=3$ &  Difference\\ 
\hline 
\hline
[0, 0, 0, 1]& [1, 0, 1, 1]& [1, 0, 0, 1, 1, 0, 0, 0, 0]& 0.29& 0.76& 	0.47\\
\hline
[0, 0, 0, 1]& [1, 0, 1, 1]& [1, 0, 0, 1, 1, 0, 0, 0, 1]&  0.29	& 0.77	& 0.48\\
\hline
[0, 0, 0, 1]&[1, 0, 1, 1]&  [1, 0, 0, 1, 1, 0, 0, 1, 0]&  0.29& 0.77	& 0.48\\
\hline
[0, 0, 0, 1]& [1, 0, 1, 1]&  [1, 0, 0, 1, 1, 0, 0, 1, 1]& 	 0.29& 	0.77	& 0.48\\
\hline
[0, 0, 0, 1]& [1, 0, 1, 1]& [1, 0, 1, 1, 1, 0, 0, 0, 0]& 0.29& 	0.76	& 0.47\\
\hline
[0, 0, 0, 1]& [1, 0, 1, 1]&  [1, 0, 1, 1, 1, 0, 0, 0, 1]& 	 0.29& 	0.75& 	0.46\\
\hline
[0, 0, 0, 1]& [1, 0, 1, 1]&  [1, 0, 1, 1, 1, 0, 0, 1, 0]& 	 0.29	& 	0.77& 	0.48\\
\hline
[0, 0, 0, 1]&  [1, 0, 1, 1]&  [1, 0, 1, 1, 1, 0, 0, 1, 1]&  0.29& 	0.76& 	0.47\\
\hline
 [0, 0, 1, 0]&  [1, 0, 1, 1]&  [1, 0, 0, 1, 1, 0, 0, 0, 0]	&   0.21	&  	0.76	&  0.55\\
\hline
 [0, 0, 1, 0]& [1, 0, 1, 1]&   [1, 0, 0, 1, 1, 0, 0, 0, 1]&  0.21	&  	0.77	&  0.56\\
\hline
 [0, 0, 1, 0] & [1, 0, 1, 1]&   [1, 0, 0, 1, 1, 0, 0, 1, 0]&  	 0.21	&  	0.77&  	0.56\\
\hline
 [0, 0, 1, 0]&[1, 0, 1, 1]&   [1, 0, 0, 1, 1, 0, 0, 1, 1]&  0.21	& 0.77	&  0.56\\
\hline
 [0, 0, 1, 0]	&  [1, 0, 1, 1]	&  [1, 0, 1, 1, 1, 0, 0, 1, 1]&  0.21		&  	0.76		&  0.55\\
\hline
 [0, 0, 1, 1]	&   [1, 0, 1, 1]	& 	 [1, 0, 0, 1, 1, 0, 0, 1, 1]	&  	 0.36	&  		0.81		&  0.45\\
\hline
 [0, 0, 1, 1]	& [1, 0, 1, 1]	&   [1, 0, 1, 1, 1, 0, 0, 1, 1]	&  	 0.36		&  	0.84		&  0.48\\
\hline
 [1, 1, 0, 0]	&  [0, 1, 0, 0]	&  [0, 1, 0, 0, 0, 1, 1, 0, 0]	 	&  0.36		&  	0.84	&  	0.48\\
\hline
 [1, 1, 0, 0]	&  [0, 1, 0, 0]	&   [0, 1, 1, 0, 0, 1, 1, 0, 0]	& 	 0.36		&  	0.81		&  0.45\\
\hline
 [1, 1, 0, 1]	&   [0, 1, 0, 0]	&  	 [0, 1, 0, 0, 0, 1, 1, 0, 0]	&  	 0.21		&  0.76	& 0.55\\
\hline
[1, 1, 0, 1]	&   [0, 1, 0, 0]	&  	 [0, 1, 0, 0, 0, 1, 1, 0, 1]	&   0.21		&  	0.77		&  0.56\\
\hline
[1, 1, 0, 1]&   [0, 1, 0, 0]	&  [0, 1, 0, 0, 0, 1, 1, 1, 0]		&   0.21		& 	0.75	&  	0.54\\
\hline
[1, 1, 0, 1]&  [0, 1, 0, 0]	&   [0, 1, 0, 0, 0, 1, 1, 1, 1]	&  	 0.21		&  	0.76	&  	0.55\\
\hline
[1, 1, 0, 1]	&  [0, 1, 0, 0]	&  [0, 1, 1, 0, 0, 1, 1, 0, 0]	 	&  0.21		&  	0.77		&  0.56\\
\hline
 [1, 1, 0, 1]	&  [0, 1, 0, 0]	&  [0, 1, 1, 0, 0, 1, 1, 0, 1]	&  	 0.21		&  	0.77	&  	0.56\\
\hline
[1, 1, 0, 1]	&  [0, 1, 0, 0]	&  [0, 1, 1, 0, 0, 1, 1, 1, 0]		&   0.21	&  		0.77	& 	0.56\\
\hline
[1, 1, 0, 1]	& [0, 1, 0, 0]	&  [0, 1, 1, 0, 0, 1, 1, 1, 1]	&   0.21	&  0.76	&  	0.55\\
\hline
[1, 1, 1, 0]	& [0, 1, 0, 0]	& [0, 1, 0, 0, 0, 1, 1, 0, 0]		&   0.29		&  	0.76	&  	0.47\\
\hline
[1, 1, 1, 0]	& [0, 1, 0, 0]	&  [0, 1, 0, 0, 0, 1, 1, 0, 1]	& 0.29		& 	0.77	& 	0.48\\
\hline
[1, 1, 1, 0]	& [0, 1, 0, 0]	& [0, 1, 0, 0, 0, 1, 1, 1, 0]		& 0.29	&		0.75		& 0.46\\
\hline
[1, 1, 1, 0]	& [0, 1, 0, 0]	&  [0, 1, 0, 0, 0, 1, 1, 1, 1]	& 0.29		&	0.76	& 	0.47\\
\hline
[1, 1, 1, 0]	& [0, 1, 0, 0]	&  [0, 1, 1, 0, 0, 1, 1, 0, 0]	& 0.29		& 	0.77	& 	0.48\\
\hline
[1, 1, 1, 0]	& [0, 1, 0, 0]	&  [0, 1, 1, 0, 0, 1, 1, 0, 1]	& 0.29		& 	0.77	& 	0.48\\
\hline
[1, 1, 1, 0]	& [0, 1, 0, 0]	&  [0, 1, 1, 0, 0, 1, 1, 1, 0]	& 0.29		& 0.77	& 	0.48\\
\hline
[1, 1, 1, 0]	&  [0, 1, 0, 0]	& 	 [0, 1, 1, 0, 0, 1, 1, 1, 1]		&  0.29		& 	0.76		& 0.47\\
\hline
\hline
\end{tabular}
}\\
\textit{\begin{tiny}
*W.P. denotes winning probability.
\end{tiny}}

\end{table}

\begin{algorithm}
\setlength{\textfloatsep}{0.05cm}
\setlength{\floatsep}{0.05cm}
\begin{large}
\KwIn{$n$ number of maximally entangled bipartite state  $\ket{\Psi_{AB}}$ in an Hilbert space $\mathbb{C}^d\times \mathbb{C}^d$ which is of the form $\sum_{j=1}^{d}\frac{1}{\sqrt{d}}\ket{j}\otimes\ket{j}$, where $\lbrace \ket{j} \rbrace$ is the standard basis of $\mathbb{C}^d$ and $d\in \lbrace 2,3 \rbrace$ is fixed but unknown. }
\KwOut{The value of $d$.}

\begin{enumerate}
\item For rounds $i \in \lbrace 1,\ldots ,n\rbrace$\\
\begin{enumerate}
  \item Referee chooses $x_i \in \lbrace 0, 1\rbrace $ and $y_i \in  \lbrace 0, 1\rbrace$ uniformly at random
  \item 
  $\bullet$ If $x_i = 0$, Alice measures the first particle of the $i$-th
entangled state in the standard basis $\lbrace \ket{0},\ket{1},\ket{2} \rbrace$\\ 
$\bullet$ If $x_i = 1$, she measures that in the Fourier basis $\lbrace \ket{0_x},\ket{1_x},\ket{2_x} \rbrace$, where \\
$\ket{0_{x}}=\dfrac{1}{\sqrt{3}}(\ket{0}+\ket{1}+\ket{2})$\\
$\ket{1_{x}}=\dfrac{1}{\sqrt{3}}(\ket{0}+\omega\ket{1} +\omega^{2}\ket{2})$\\
$\ket{2_{x}}=\dfrac{1}{\sqrt{3}}(\ket{0}+\omega^{2}\ket{1} +\omega\ket{2})$
\\
and $\omega=e^{{2\pi i}/{3}}$ 
(if $d=2$, it will be the Hadamard basis)
  \item Similarly,\\
  $\bullet$ If $y_i = 0$, Bob measures the second particle of the entangled state in $ \lbrace\ket{\psi_0},\ket{\psi_1},\ket{\psi_2}\rbrace$ basis , where\\
$\ket{\psi_0} = \cos \theta_0 \ket{0} + \sin \theta_0  \cos \theta_1 \ket{1} + \sin \theta_0 \sin \theta_1 \ket{2}$\\
$\ket{\psi_1} = \sin \theta_0 \ket{0} - \cos \theta_0  \cos \theta_1 \ket{1} - \cos \theta_0 \sin \theta_1 \ket{2}$ \\
$\ket{\psi_2} = \sin \theta_1 \ket{1} +  \cos \theta_1 \ket{2}$\\
$0\leqslant \theta_0,\theta_1\leqslant 2\pi$

$\bullet$ If $y_i = 1$, he measures that in $ \lbrace\ket{\phi_0},\ket{\phi_1},\ket{\phi_2}\rbrace$ basis,  where\\
$\ket{\phi_0} = \cos \theta_1 \ket{0} + \sin \theta_1 \cos \theta_0 \ket{1} + \sin \theta_1 \sin \theta_0 \ket{2}$\\
$\ket{\phi_1} = \sin \theta_1 \ket{0} - \cos \theta_1 \cos \theta_0 \ket{1} - \cos \theta_1 \sin \theta_0 \ket{2}$ \\
$\ket{\phi_2} = \sin \theta_0 \ket{1} +  \cos \theta_0 \ket{2}$\\ 
$0\leqslant \theta_0,\theta_1\leqslant 2\pi$
  
  \item The output is recorded as $a_i (b_i ) \in \lbrace 0, 1,2\rbrace$ for the
first (second) particle. The encoding for $a_i (b_i )$ is as
follows:\\
$\bullet$ For the first particle of each pair, $a_i = j$ if the
measurement result is $\ket{j}$ or $\ket{j_x}$\\
  $\bullet$ For the second particle of each pair,\\
   $b_i = 0$ if the measurement result is $\ket{\psi_0}$ or $\ket{\phi_0}$\\
   $b_i=1$ if the measurement result is $\ket{\psi_1}$ or $\ket{\phi_1}$\\
   $b_i=2$ if the measurement result is $\ket{\psi_2}$ or $\ket{\phi_2}$

\item For the test round $i$, define \\

\begin{equation*}
	Y_i=
		\begin{cases}
			1 & \text{ if $x_i\wedge y_i=g(a_i,b_i)$ where g=Embedded~XOR}\\
			0 & \text{ if otherwise}
		\end{cases}
\end{equation*}

\end{enumerate}
\item Referee calculates $S=\dfrac{1}{n}\sum Y_i$
\item If $S\approx 0.85$, return $d=2$
and if $S\approx 0.76$, return $d=3$

\end{enumerate}

\end{large}
\caption{Dimension distinguisher of maximally entangled state}
\label{algo:dim_dist}
\setlength{\textfloatsep}{0.05cm}
\setlength{\floatsep}{0.05cm}
\end{algorithm}

\section{Conclusion}
Dimensionality of the states act as a resource in quantum information processing tasks. For many protocols, the performance as well as security depends on the particular value of the dimension. For this reason, dimensionality testing is very important. There have been several works on dimension witness. We take a different route by constructing dimension distinguishers based on our generalized version of the the CHSH game. We demonstrate several classes of practical distinguishers between 2 and 3 dimensions.




\end{document}